\RequirePackage{fix-cm}
\documentclass[smallextended]{svjour3}       
\smartqed  
\usepackage{graphicx}
\usepackage{epsfig}
\begin{document}

\title{Area(or Entropy) Product Formula for a Regular Black Hole 
}
\author{Parthapratim Pradhan}



\institute{    \at
               Department of Physics\\
               Vivekananda Satavarshiki Mahavidyalaya\\
               (Affiliated to Vidyasagar University)\\
               Manikpara, Jhargram, West Midnapur\\
               West Bengal~721513, India \\
               \email{pppadhan77@gmail.com}
          }

\date{Received: date / Accepted: date}

\maketitle

\begin{abstract}
We compute the area(or entropy) product formula for a regular black hole derived by
Ay\'{o}n-Beato and Garc\'{i}a in 1998\cite{abg}. By explicit and exact
calculation, it is shown that the entropy product formula of two physical horizons 
strictly \emph{depends} upon the ADM mass parameter that means it is \emph{not} an 
universal(mass-independent) quantity. But a slightly more complicated function of event horizon 
area and Cauchy horizon area is indeed a \emph{mass-independent} quantity.  
We also compute other thermodynamic properties of the said black hole. 
We further study the stability of such black hole by computing the specific 
heat for both the horizons.  It has been observed  that under certain condition 
the black hole possesses second order phase transition. The pictorial diagram of the 
phase transition is given.
\end{abstract}

\section{Introduction}
The area (or entropy ) products of event horizon and Cauchy horizons have found an intriguing 
field of research in the recent years both in  general relativity community\cite{ah09}  and 
String/M-theory community\cite{cgp11} (see also\cite{cr12,sd12,pp14}). Particularly, the interest 
involved in the relations that are independent of the black hole(BH) mass or ADM(Arnowitt-Deser-Misner)
mass. It has been  explicitly examined for the Kerr BH as well as Kerr-Newman(KN) BH. 

For KN BH, it has been shown that the area product formula of inner
horizon(${\cal H}^{-}$) and outer horizons(${\cal H}^{+}$) should read\cite{ah09}
\begin{eqnarray}
{\cal A}_{-}{\cal A}_{+} &=& 64 \pi^2 J^2+16\pi^2Q^4 ~.\label{prKN}
\end{eqnarray}
where $J$ and $Q$ are the angular momentum and charge respectively.
 
For Kerr BH, this product formula reduces to the following form:
\begin{eqnarray}
{\cal A}_{-}{\cal A}_{+}  &=& 64 \pi^2 J^2 ~.\label{prK}
\end{eqnarray}
It implies that these are the relations that only involve quasi-local quantities defined at the horizons
(areas, charges, angular momenta etc.). Therefore, these equations are particularly interesting as they could 
hold in more general situations where the BHs are perturbed by surrounding matter\cite{ah09}. The area product 
for the Kerr BH is of a \emph{universal} in nature because it holds independently of the environment of the BH.

The motivation comes from the idea of Visser's\cite{mv13} work. Where the author has been shown by 
explicit computations that the area product is not mass independent for Reissner Nordstr{\o}m-AdS BH.
But slightly more complicated function of physical horizon areas that are mass-independent. Similarly, it has 
been shown explicitly by Hennig\cite{jh} that there exists a mass-independent area product relation for KN-AdS 
space-time. 

In this work, we wish to examine the area product formula for a regular BH in general relativity coupled to a 
nonlinear electromagnetic field. This exact BH solution was first described by Ay\'{o}n-Beato and Garc\'{i}a(ABG) 
in 1998\cite{abg}. This is a curvature singularity free solution in Einstein's general relativity. We first 
calculate the area product in terms of two physical horizons namely, event horizon and Cauchy horizon as 
described by ABG in his work\cite{abg}. We show that in this case the area(or entropy) product is 
\emph{ not mass independent}. Next we evaluate the explicit formula for areas by solving the quartic 
horizon equation. We show that a slightly more complicated combination of two physical horizons that 
is \emph{mass independent}(universal). 

The term ``universal'' is used throughout the paper for relations that are mass-independent. However,  this 
term has traditionally been used if a formula holds in a very wide class of 
space-times[e.g. Kerr space-time(in this case a simple product is sufficient)] and 
not just in a rather small family of exact solutions. Hence it is more appropriate to say 
that the mass-independent relation we found could \emph{turn out to be universal}, i.e. hold 
in a larger context.   

The plan of the paper is as follows. In Sec. 2, we described the basic properties of the ABG BH and 
computed various thermodynamic properties.  Finally, we conclude our discussions in Sec. 3.

\section{ABG BH:}
In 1998, Ay\'{o}n-Beato and Garc\'{i}a\cite{abg} derived a singularity-free solution of the Einstein 
field equations coupled to a non-linear electrodynamics described by the metric:
\begin{eqnarray}
ds^2=-{\cal G}(r)dt^{2}+\frac{dr^{2}}{{\cal G}(r)}+r^{2}(d\theta^{2}+\sin^{2}\theta d\phi^{2}) ~.\label{abg1}
\end{eqnarray}
where the function ${\cal G}(r)$ is defined by
\begin{eqnarray}
{\cal G}(r) &=& 1-\frac{2mr^2}{(r^2+q^2)^{\frac{3}{2}}}+\frac{q^2r^2}{(r^2+q^2)^2} ~.\label{abg2}
\end{eqnarray}
where $m$ is the mass of the BH and $q$ is the monopole charge.

This is a regular exact black hole solution in general relativity. The
source is a nonlinear electrodynamic field satisfying the weak energy condition, which is in the
limit of weak field becomes the Maxwell field. The metric as well as the curvature 
invariants $R, R_{ab}R^{ab}, R_{abcd}R^{abcd}$ and the electric field are regular 
everywhere in the space-time\cite{ansoldi}. Hence in this sense it is called a regular BH in the 
Einstein-Maxwell gravity. The first regular BH model was developed by Bardeen in 1968
\cite{bard}.

The fact is that Smarr's mass formula and Bose-Dadhich identity(Difference between Brown-York 
quasi-local energy and Komar\cite{ak59} charge on the horizon) do not hold for this regular BH 
when we have taken into account the nonlinear electrodynamics\cite{breton,bose,balrat}, whereas 
the first law of BH thermodynamics does hold.

It may also be noted that the metric function  asymptotically behaves as a
Reissner Nordstr{\o}m space-time i.e.,
\begin{eqnarray}
{\cal G}(r) \sim 1-\frac{2m}{r}+\frac{q^2}{r^2} ~.\label{abg3}
\end{eqnarray}

The BH horizons as derived in\cite{abg}:
\begin{eqnarray} 
r_{\pm} &=& |q|\sqrt{\left[\frac{1}{4\alpha} + \frac{\sqrt{Z}}{12\alpha} \pm\frac{\sqrt{6}}{12\alpha}
\sqrt{\frac{9}{2}-12\alpha^2-\frac{Z}{6}-\frac{9(12\alpha^2-1)}{\sqrt{Z}}} \right]^2-1}. ~\label{abg4}\\
Z &=& 6 \left[\frac{3}{2}-4\alpha^2+\alpha Z_{1}^{\frac{1}{3}}-\frac{4\alpha(11\alpha^2-3)}
{Z_{1}^{\frac{1}{3}}} \right] . ~\label{abg5}\\
Z_{1} &=& 4\left[9\alpha+74\alpha^3+\sqrt{27(400\alpha^6-112\alpha^4+47\alpha^2-4)}\right] 
. ~\label{abg6} 
\end{eqnarray}
where $\alpha=\frac{|q|}{2m}$. As is $r_{+}$ corresponds to event horizon and  
$r_{-}$ corresponds to Cauchy horizon. 

The area of this BH is given by
\begin{eqnarray}
{\cal A}_{\pm} &=& \int^{2\pi}_0\int^\pi_0 \sqrt{g_{\theta\theta} g_{\phi \phi}}d\theta d\phi
=4 \pi r_{\pm}^2 \\
               &=& 4 \pi q^2\left[\frac{1}{4\alpha^2}-\frac{3}{2}-\frac{3(12\alpha^2-1)}{8\alpha^2\sqrt{Z}} 
                   +\frac{\sqrt{Z}}{24\alpha^2}\pm \frac{(3+\sqrt{Z})}{12\sqrt{6}\alpha^2}\sqrt{\frac{9}{2}-12\alpha^2
                   -\frac{Z}{6}-\frac{9(12\alpha^2-1)}{\sqrt{Z}}}\right]
                  ~. \nonumber
\end{eqnarray}
and their product yields
\begin{eqnarray}
{\cal A}_{+}{\cal A}_{-} &=&  16\pi^2 q^4  \times  \label{abg8} 
\end{eqnarray}
$$
\left[\left(\frac{1}{4\alpha^2}-\frac{3}{2}-\frac{3(12\alpha^2-1)}{8\alpha^2\sqrt{Z}} 
+\frac{\sqrt{Z}}{24\alpha^2}\right)^2-\frac{(3+\sqrt{Z})^2}{864\alpha^4} 
\left(\frac{9}{2}-12\alpha^2-\frac{Z}{6}-\frac{9(12\alpha^2-1)}{\sqrt{Z}}\right)\right] 
$$
From the above expression it is clearly evident  that the product strictly depends upon the
ADM mass parameter. Thus the area product of all the horizons have no nice quantization features.
Nor does it have any universal features. 

In the following subsection we shall see a more complicated function of horizon areas turned 
out to be mass-independent.

The BH entropy\cite{bk73} computed at ${\cal H}^{\pm}$ 
\begin{eqnarray}
{\cal S}_{\pm} &= & \pi r_{\pm}^2 .~\label{abg9}
\end{eqnarray}
Consequently, the entropy product formula is given by
\begin{eqnarray}
{\cal S}_+{\cal S}_- &=&  \pi^2 q^4 \times \label{10}
\end{eqnarray}
$$
\left[\left(\frac{1}{4\alpha^2}-\frac{3}{2}-
\frac{3(12\alpha^2-1)}{8\alpha^2\sqrt{Z}} +\frac{\sqrt{Z}}{24\alpha^2}\right)^2-\frac{(3+\sqrt{Z})^2}{864\alpha^4} 
\left(\frac{9}{2}-12\alpha^2-\frac{Z}{6}-\frac{9(12\alpha^2-1)}{\sqrt{Z}}\right)\right] .~\label{abg10}
$$
It implies that the entropy product formula does depend on the ADM mass parameter, thus 
it should not be a mass-independent quantity. 

The Hawking\cite{bcw73} temperature on ${\cal H}^{\pm}$ reads off
\begin{eqnarray}
T_{\pm} &=& \frac{\kappa_{\pm}}{2\pi}=\frac{1}{4\pi}
\left[\frac{2mr_{\pm}(r_{\pm}^2-2q^2)}{(r_{\pm}^2+q^2)^{\frac{5}{2}}}-
\frac{2q^2r_{\pm}(r_{\pm}^2-q^2)}{(r_{\pm}^2+q^2)^3} \right] .~\label{abg11}
\end{eqnarray}


It may be noted that the Hawking temperature product is
depend on the mass parameter and thus it is not an universal quantity.
\subsection{Exact results:}
From Eq. (\ref{abg2}), we obtain BH horizon equation by setting the condition ${\cal G}(r)=0$:
\begin{eqnarray}
r^{8}+(6q^2-4m^2)r^{6}+(11q^4-4m^2q^2)r^{4} +6q^6r^2+q^8 = 0 ~.\label{hab}
\end{eqnarray}
Let us put $r^2=x$, we get the quartic equation:
\begin{eqnarray}
x^{4}+(6q^2-4m^2)x^{3}+(11q^4-4m^2q^2)x^{2} +6q^6x+q^8 = 0 ~.\label{hab1}
\end{eqnarray}
Using Vieta's theorem, we find the four equations:
\begin{eqnarray}
x_{1}+x_{2}+x_{3}+x_{4} &=& 4m^2-6q^2 ~.\label{eq1}\\
x_{1}x_{2}+x_{1}x_{3}+x_{1}x_{4}+x_{2}x_{3}+x_{2}x_{4}+x_{3}x_{4} &=& 11q^4- 4m^2q^2 ~.\label{eq2}\\
x_{1}x_{2}x_{3}+x_{2}x_{3}x_{4}+x_{3}x_{4}x_{1}+x_{2}x_{4}x_{1} &=& -6q^2 ~.\label{eq3}\\
x_{1}x_{2}x_{3}x_{4} &=& q^8 ~.\label{eq4}
\end{eqnarray}
From Eqs. (\ref{eq1},\ref{eq2}), eliminating $m$, we get the following equation:
$$
x_{1}x_{2}+x_{1}x_{3}+x_{1}x_{4}+x_{2}x_{3}+x_{2}x_{4}+x_{3}x_{4}+
$$
\begin{eqnarray}
q^2(x_{1}+x_{2}+x_{3}+x_{4}) &=& 5q^4 
~.\label{eq5}
\end{eqnarray}
In terms of area ${\cal A}_{i}=4\pi r_{i}^2=4 \pi x_{i}$($i=1,2,3,4$), we obtain mass independent 
relations in terms of four horizon areas:
$$
{\cal A}_{1}{\cal A}_{2}+{\cal A}_{1}{\cal A}_{3}+{\cal A}_{1}{\cal A}_{4}+{\cal A}_{2}{\cal A}_{3}+{\cal A}_{2}{\cal A}_{4}
+{\cal A}_{3}{\cal A}_{4}+
$$
\begin{eqnarray}
4\pi q^2({\cal A}_{1}+{\cal A}_{2}+{\cal A}_{3}+{\cal A}_{4}) &=& 80 \pi^2 q^4  ~.\label{eq6}
\end{eqnarray}
\begin{eqnarray}
{\cal A}_{1}{\cal A}_{2}{\cal A}_{3}+{\cal A}_{2}{\cal A}_{3}{\cal A}_{4}+{\cal A}_{3}{\cal A}_{4}{\cal A}_{1}
+ {\cal A}_{2}{\cal A}_{4}{\cal A}_{1} &=& -384\pi^3 q^6  ~.\label{eq7}
\end{eqnarray}
and 
\begin{eqnarray}
{\cal A}_{1}{\cal A}_{2}{\cal A}_{3}{\cal A}_{4} &=& 256 \pi^4 q^8  ~.\label{eq8}
\end{eqnarray}
These are exact mass-independent and charge dependent relations for ABG BH. Now we turn to the case 
of area(or entropy) product in terms of two physical BH areas.
The quartic equation(\ref{hab1}) has at least two real roots corresponding to event horizon and Cauchy horizon.
Now eliminating third and fourth roots, we get the following equations:
\begin{eqnarray}
 \left[x_{1}x_{2}(x_{1}+x_{2})+(6q^2-4m^2)x_{1}x_{2}-q^8\frac{(x_{1}+x_{2})}{x_{1}x_{2}}\right] 
 &=& 6q^6  ~.\label{eq9}
\end{eqnarray}
and
\begin{eqnarray}
 \frac{1}{x_{1}+x_{2}+q^2}\left[(x_{1}+x_{2})^2+6q^2(x_{1}+x_{2})-x_{1}x_{2}-\frac{q^8}{x_{1}x_{2}}+11q^4\right] 
 &=& 4m^2  ~.\label{eq10}
\end{eqnarray}
From these equations, we eliminate the mass parameter, we obtain the following single mass independent 
equation:
$$
x_{1}x_{2}(x_{1}+x_{2})+6q^2x_{1}x_{2}-q^8\frac{(x_{1}+x_{2})}{x_{1}x_{2}}-
$$
\begin{eqnarray}
 \frac{x_{1}x_{2}}{x_{1}+x_{2}+q^2}\left[(x_{1}+x_{2})^2+6q^2(x_{1}+x_{2})-x_{1}x_{2}-\frac{q^8}{x_{1}x_{2}}+11q^4\right] 
 &=& 6q^6  ~.\label{eq11}
\end{eqnarray}
In terms of two BH horizons area, we get the following mass independent relation:
$$
{\cal A}_{1}{\cal A}_{2}({\cal A}_{1}+{\cal A}_{2})+24 \pi q^2 {\cal A}_{1}{\cal A}_{2}-
256 \pi ^4 q^8 (\frac{{\cal A}_{1}+{\cal A}_{2}}{{\cal A}_{1}{\cal A}_{2}})-
$$
$$
\frac{{\cal A}_{1}{\cal A}_{2}}{{\cal A}_{1}+{\cal A}_{2} +4\pi q^2}\left[ ({\cal A}_{1}+{\cal A}_{2})^2 
+24 \pi q^2 ({\cal A}_{1}+{\cal A}_{2})-{\cal A}_{1}{\cal A}_{2}-\frac{256 \pi^4 q^8}
{{\cal A}_{1}{\cal A}_{2}} +176 \pi^2 q^4 \right]
$$
\begin{eqnarray}
 &=& 384 \pi^3 q^6  ~.\label{eq12}
\end{eqnarray}
It seems that it is a more complicated function of event and Cauchy horizon area, but it is at least 
mass-independent. So it is not straightforward as a simple product of horizon area of ${\cal H}^{\pm}$ as we have seen in Eq.(\ref{abg8}). This is the key observation of our work.

From these relations, we can easily obtain the entropy functional relations
and irreducible mass ${m}_{irr}$ product relations by substituting 
${\cal A}_{1} = 16 \pi ({m}_{irr, 1})^2$ and ${\cal A}_{2} = 16 \pi ({m}_{irr,2})^2$. Note that 
(+,-) and (1,2) are synonymous for outer and inner horizons. It may be noted that the mass independent 
functions of event and Cauchy horizon areas in spherically symmetry space-time is closely related to the 
quasi-local mass $m(r)$, which is a Laurent polynomial\cite{mv13} of the aerial radius 
$r=\sqrt{\frac{{\cal A}}{4\pi}}=\sqrt{\frac{{\cal S}}{\pi}}$.

\subsection{Heat Capacity $C_{\pm}$ on ${\cal H}^{\pm}$:}
 The specific heat on ${\cal H}^{\pm}$ is defined by
\begin{eqnarray}
C_{\pm} &=& \frac{\partial{m}}{\partial T_{\pm}} .~\label{abg17}
\end{eqnarray}
which is an important parameter to study the thermodynamic stability of a BH. 
For ABG BH, we find
\begin{eqnarray}
C_{\pm} &=&  \frac{2\pi (r_{\pm}^2+q^2)^{\frac{5}{2}}\left[q^2r_{\pm}^4-(r_{\pm}^2-2q^2)(r_{\pm}^2+q^2)^{2}\right]}
{r_{\pm}\left[(r_{\pm}^4-7q^2r_{\pm}^2-2q^4)(r_{\pm}^2+q^2)^{2}-3q^2r_{\pm}^4(r_{\pm}^2-q^2)\right]} 
.~\label{abg18}
\end{eqnarray}
Let us analyze the above expression of specific heat for a different regime in the parameter space.

Case I: The specific  heat $C_{\pm}$ is positive when $q^2r_{\pm}^4>(r_{\pm}^2-2q^2)(r_{\pm}^2+q^2)
^{2}$ and $(r_{\pm}^4-7q^2r_{\pm}^2-2q^4)(r_{\pm}^2+q^2)^{2}>3q^2r_{\pm}^4(r_{\pm}^2-q^2)$, in this case 
the BH is thermodynamically stable.

Case II: The  specific  heat $C_{\pm}$ is negative when $q^2r_{\pm}^4<(r_{\pm}^2-2q^2)(r_{\pm}^2+q^2)
^{2}$ and $(r_{\pm}^4-7q^2r_{\pm}^2-2q^4)(r_{\pm}^2+q^2)^{2}>3q^2r_{\pm}^4(r_{\pm}^2-q^2)$ or 
$q^2r_{\pm}^4>(r_{\pm}^2-2q^2)(r_{\pm}^2+q^2)^{2}$ and 
$(r_{\pm}^4-7q^2r_{\pm}^2-2q^4)(r_{\pm}^2+q^2)^{2}<3q^2r_{\pm}^4(r_{\pm}^2-q^2)$, in this case 
the BH is thermodynamically unstable.

Case III:  The  specific  heat $C_{\pm}$ blows up when 
$(r_{\pm}^4-7q^2r_{\pm}^2-2q^4)(r_{\pm}^2+q^2)^{2}=3q^2r_{\pm}^4(r_{\pm}^2-q^2)$, in this case 
the BH undergoes a second order phase transition.

It should be noted that the product of specific heat on ${\cal H}^{\pm}$ depends on mass 
parameter and charge parameter. Thus the product of specific heat of both the horizons is 
not an universal quantity .

\begin{center}
\scalebox{0.4}{\includegraphics{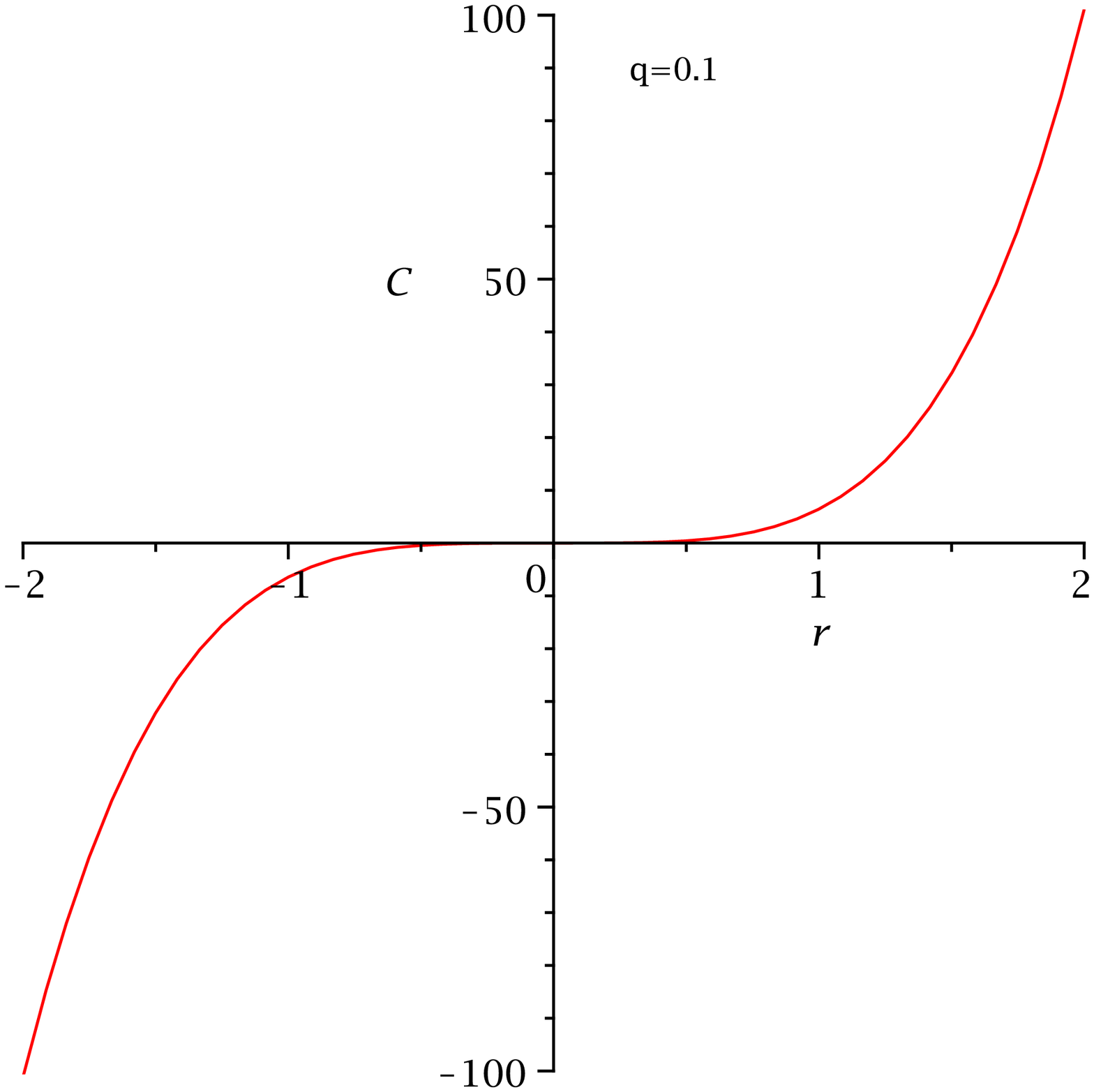}}\\
\vspace{0.02cm}
\end{center}
Figure 1. The figure shows the variation  of $C$  with $r$ for ABG BH.\\
\begin{center}
\scalebox{0.4}{\includegraphics{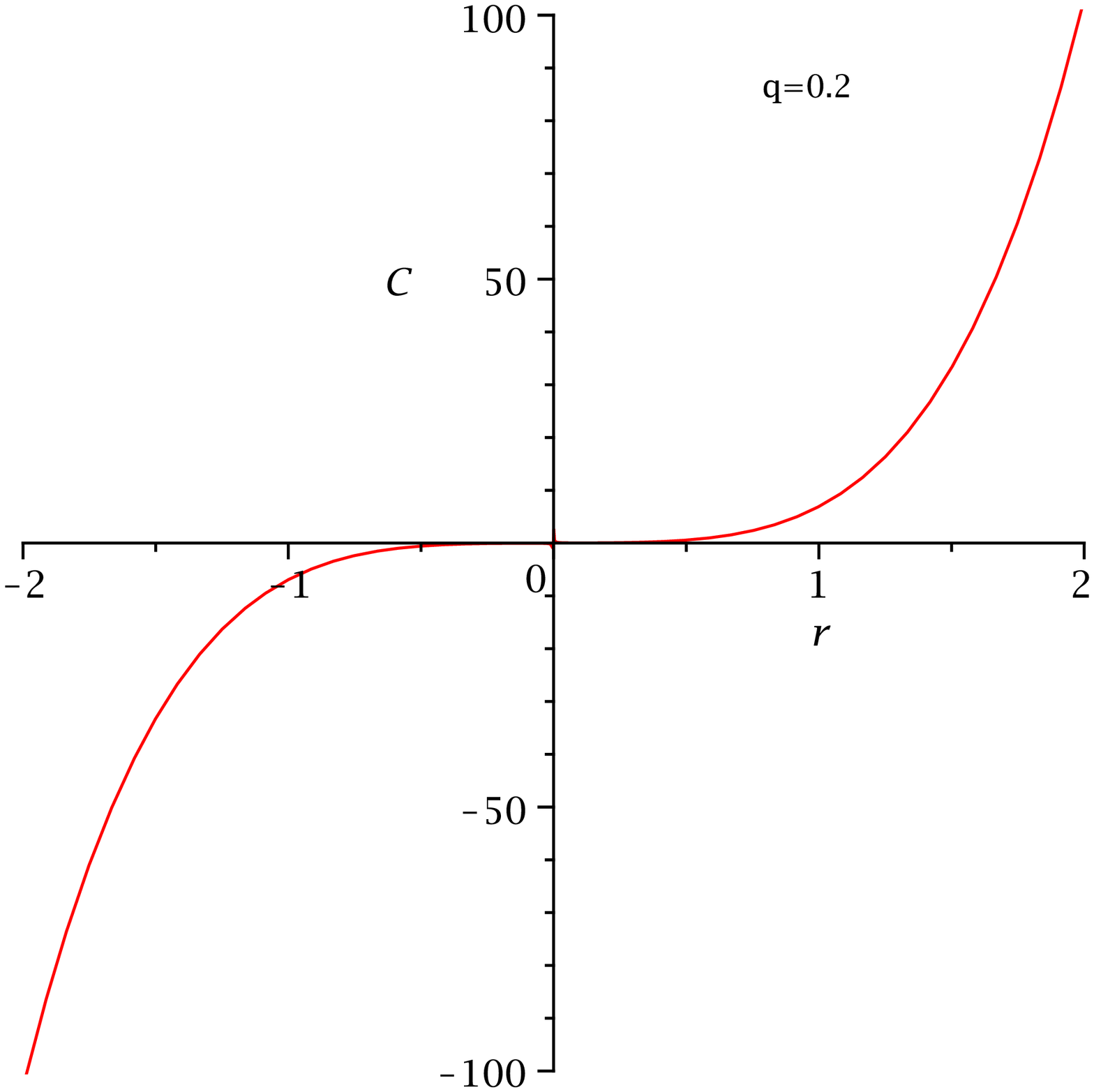}}\\
\vspace{0.02cm}
\end{center}
Figure 2. The figure shows the variation  of $C$  with $r$ for ABG BH.\\

\begin{center}
\scalebox{0.4}{\includegraphics{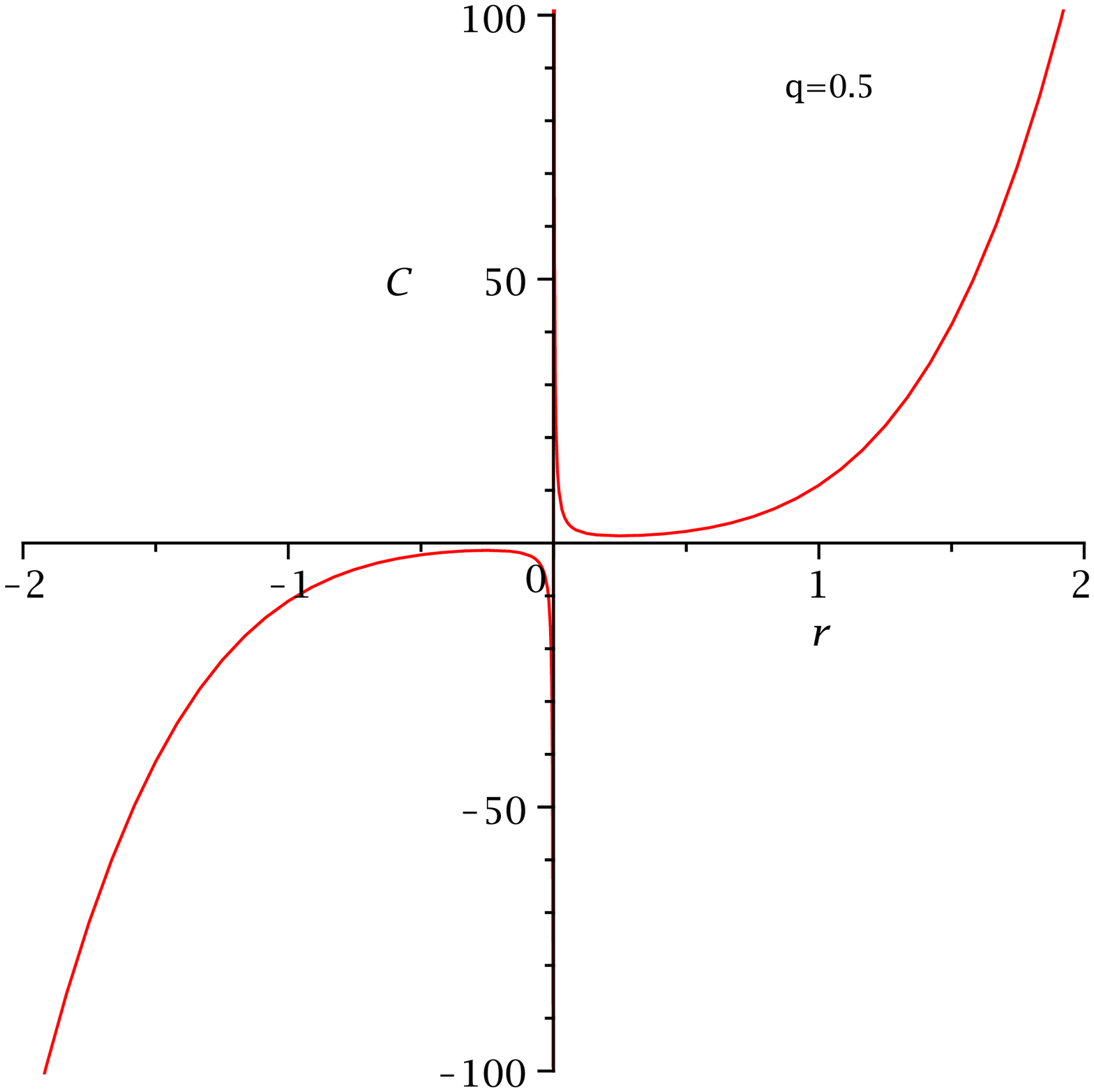}}\\
\vspace{0.02cm}
\end{center}
Figure 3. The figure shows the variation  of $C$  with $r$ for ABG BH.\\

\begin{center}
\scalebox{0.4}{\includegraphics{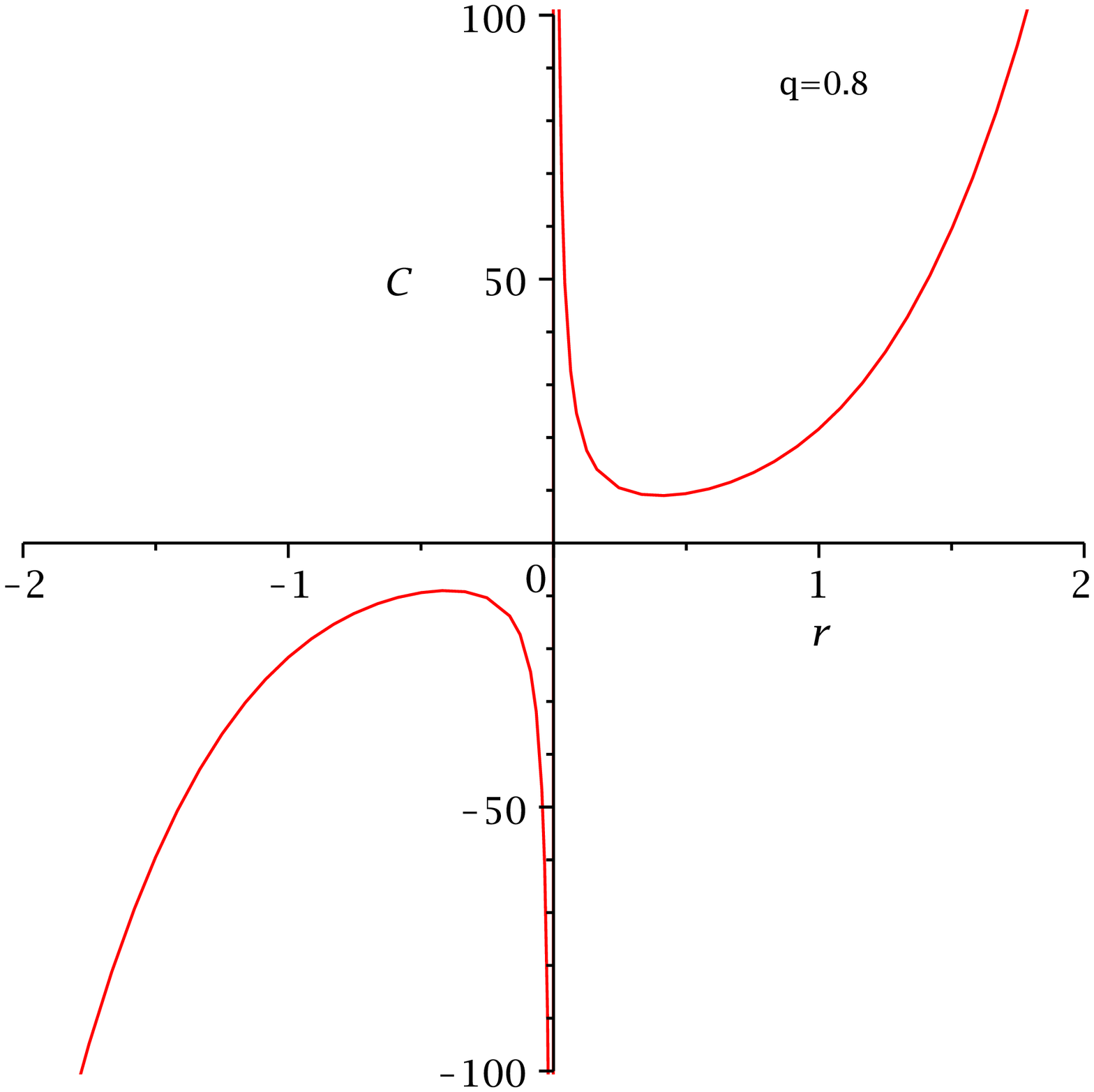}}\\
\vspace{0.02cm}
\end{center}
Figure 4. The figure shows the variation  of $C$  with $r$ for ABG BH.\\

\begin{center}
\scalebox{0.4}{\includegraphics{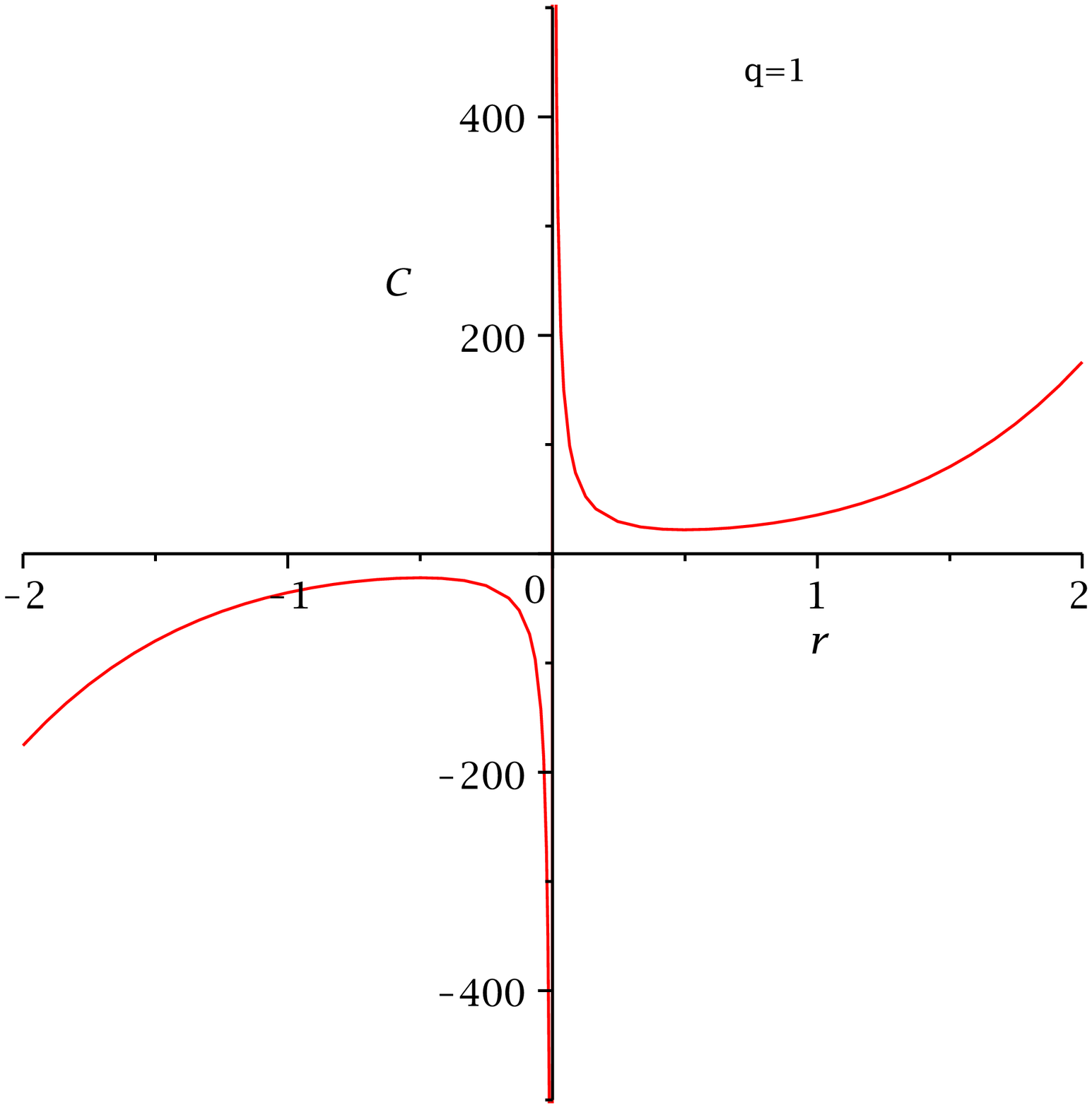}}\\
\vspace{0.02cm}
\end{center}
Figure 5. The figure shows the variation  of $C$  with $r$ for ABG BH.\\

\begin{center}
\scalebox{0.4}{\includegraphics{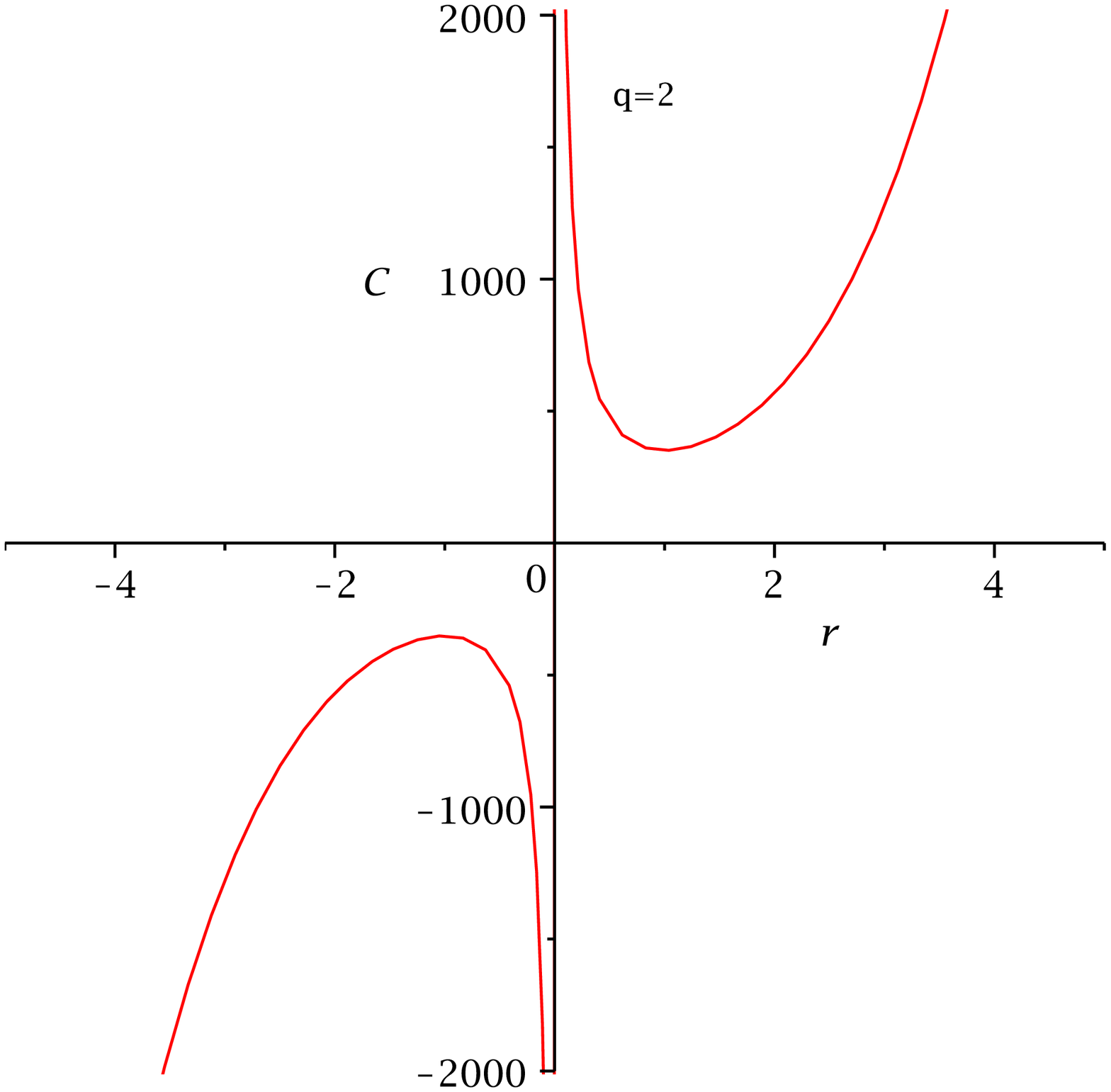}}\\
\vspace{0.02cm}
\end{center}
Figure 6. The figure shows the variation  of $C$  with $r$ for ABG BH.\\

\begin{center}
\scalebox{0.4}{\includegraphics{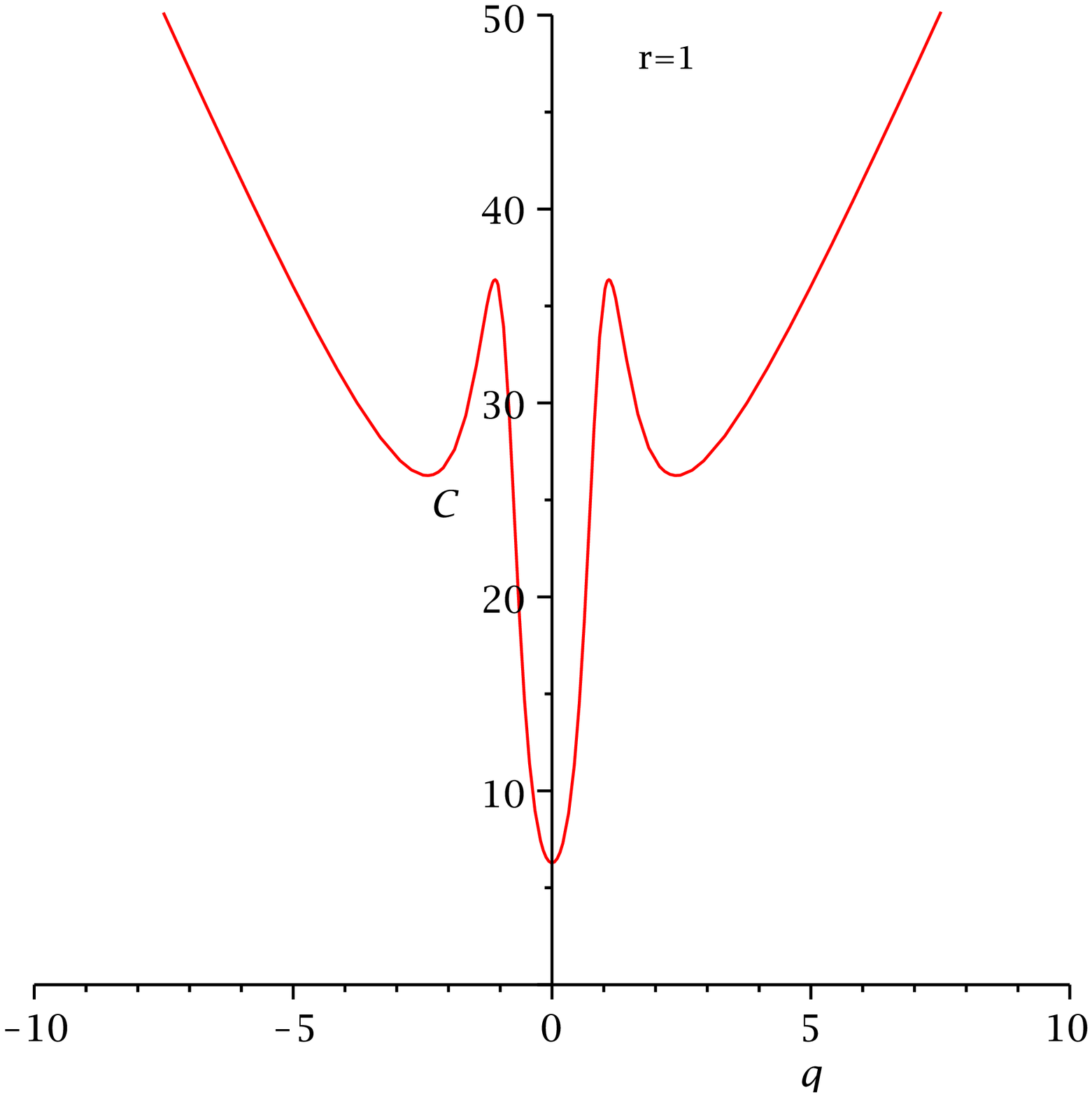}}\\
\vspace{0.02cm}
\end{center}
Figure 7. The figure shows the variation  of $C$  with $q$ for ABG BH.\\

From Fig.1- Fig. 6, we have plotted the specific heat with the horizon radius for 
various values of $q$. It follows from the diagram how BH changes its phase from stability 
region to instability region.

It should be noted that the product of specific heat on ${\cal H}^{\pm}$ depends on mass 
parameter and charge parameter. Thus the product of specific heat of both the horizons is 
not an universal quantity.

\section{Discussion:}
In this work, we studied the thermodynamic properties of a regular BH(i.e. a solution without any curvature singularities)
whose source is a nonlinear electromagnetic field. We computed various thermodynamic product formulas for this BH. We found 
that the area product of ${\cal H}^{\pm}$ explicitly depend of the mass of the BH but a more complicated function of the 
horizon areas is indeed independent of the BH mass parameter.  

Finally, we studied the stability of such BH by computing the specific heat for both the horizons. 
It has been shown that under certain condition the black hole possesses  second order phase transition.  
In conclusion, these mass independent area(or entropy) product formula might be helpful for further understanding the 
microscopic nature of BH entropy both interior and exterior which is one of the main issues in quantum gravity.

\section*{Acknowledgments}
I am really thankful to the anonymous referee for interesting suggestions.

\bibliographystyle{model1-num-names}

\end{document}